# Tuneable Drug-loading Capability of Chitosan Hydrogels with Varied Network Architectures

Giuseppe Tronci,[a, b] Hiroharu Ajiro,[c] Stephen J. Russell,[b] David J. Wood,[a] Mitsuru Akashi[c*]

[a] Biomaterials and Tissue Engineering Research Group, School of Dentistry, University of Leeds, Clarendon Way, Leeds, LS2 9LU, United Kingdom

[b] Nonwovens Research Group, Centre for Technical Textiles, School of Design, University of Leeds, Leeds, LS2 9JT, United Kingdom

[c] Department of Applied Chemistry, Graduate School of Engineering, Osaka University, 2-1, Yamadaoka, Suita, Osaka 565-0871, Japan

## Abstract

Advanced bioactive systems with defined macroscopic properties and spatio-temporal sequestration of extracellular biomacromolecules are highly desirable for next generation therapeutics. Here, chitosan hydrogels were prepared with neutral or negatively-charged crosslinkers in order to promote selective electrostatic complexation with charged drugs. Chitosan (CT) was functionalised with varied dicarboxylic acids, such as tartaric acid (TA), poly(ethylene glycol) bis(carboxymethyl) ether (PEG), 1.4-Phenylenediacetic acid (4Ph) and 5-Sulfoisophthalic acid monosodium salt (PhS), whereby PhS was hypothesised to act as a simple mimetic of heparin. ATR FT-IR showed the presence of C=O amide I, N-H amide II and C=O ester bands, providing evidence of covalent network formation. The crosslinker content was reversely quantified by $^{1}$H-NMR on partially-degraded network oligomers, so that 18 mol.-% PhS was exemplarily determined. Swellability (*SR*: 299±65–1054±121 wt.-%), compressability (*E*: 2.1±0.9–9.2±2.3 kPa), material morphology, and drug-loading capability were successfully adjusted based on the selected network architecture. Here, hydrogel incubation with model drugs of varied electrostatic charge, i.e. allura red (*AR, --*), methyl orange (*MO, -*) or methylene blue (*MB, +*), resulted in direct hydrogel-dye electrostatic complexation. Importantly, the cationic compound, *MB*, showed different incorporation behaviours, depending on the electrostatic character of the selected crosslinker. In light of this tuneable drug-loading capability, these CT hydrogels would be highly attractive as drug reservoirs towards e.g. the fabrication of tissue models *in vitro*.

---

[*] Corresponding author: Mitsuru Akashi (akashi@chem.eng.osaka-u.ac.jp)

## Graphical abstract

Systematic chitosan functionalisation resulted in hydrogels with varied network architecture, electrostatic charge and macroscopic properties, so that bespoke hydrogel loading was successfully accomplished by controlling the electrostatic dye-hydrogel complexation.

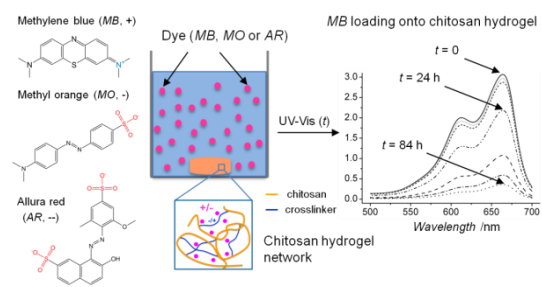

## 1. Introduction

In *in vivo* tissue engineering, multifunctional material systems should temporally mimic natural tissues, exhibiting controlled macroscopic properties and inducing specific biological signals to local stem cells [1]. *In vivo*, these functions are provided by the extracellular matrix (ECM), a highly organised supramolecular hydrogel structure binding and stabilising growth factors in a spatio-temporal, controlled manner [2]. ECM sulfated glycosaminoglycans, such as heparin, protect bound growth factors from proteolytic degradation, potentiating their bioactivity by facilitating cell receptor interactions [3]. Consequently, the design of advanced bioactive systems mimicking ECM growth factor features in a defined, application-dependent, temporal fashion, is currently considered one of the greatest challenges in regenerative medicine [4, 5].

Functional biomaterials, including hydrogels [5, 6, 7, 8, 9, 10, 11], sponges [12], scaffolds [13], electrospun meshes [14] and capsules [15], have been extensively investigated for the localised loading and controlled release of growth factors. Among the different carrier systems, hydrogels are attractive material candidates since they are generally biocompatible, biodegradable and can mimic ECM architecture over different length scales [16, 17]. At the same time, their high swelling and hard to control elasticity and degradability must be carefully addressed in order to ensure a timely and sustained material performance without compromising loading and release profiles [9]. Using functionalised dextran hydrogels, Schillemans et al. applied reversible electrostatic interactions for post-loading and release of proteins [18]. Here, protein incorporation was not homogeneous and nearly-complete release was observed following 50 hours. Aiming at sustained and localised delivery, Jeon et al. synthesised growth factor-encapsulating hydrogels [19]. Here, covalently-linked heparin segments were expected to mimic ECM growth factor-binding sites, although release was found to be mainly driven by diffusion rather than network degradation. In an effort to control loading and release capability independently of network characteristics, Freudenberg et al. described a modular system of biohybrid hydrogels based on covalently cross-linked heparin and star-shaped poly(ethylene glycols) [5]. Although mechanical properties and biofunctionality could be ruled independently, up to only about 5 wt.-% loading was observed[†], while nearly 20 wt.-% of loaded growth factor was released within 7 days. Besides biopolymer-based hydrogels, surface-modified synthetic hydrogel networks have also been established, whereby pH-triggered, hour-scale, release of model drugs was demonstrated [11]. From all the aforementioned examples, it appears rather clear that successful clinical translation of current hydrogel systems as controlled drug delivery systems has been deterred by setbacks such as (i) poor loading efficiency, (ii) large initial burst release and (iii) limited control in material properties. An interesting approach to address these challenges is based on the design of functional hydrogel reservoirs allowing for the controlled loading of a wide range of biomolecules, whereby stimulus-triggered drug release can be obtained. To reach this goal, the molecular network architecture must be thoroughly investigated, so that defined structure-property-function relationships can be established [20].

Chitosan (CT) is the second most abundant natural polymer on earth and serves as a structural polysaccharide for many phyla of lower plants and animals. CT is derived from the partial deacetylation of chitin, thereby resulting in the only linear cationic polysaccharide. It contains glucosamine and *N*-acetylglucosamine units along its polymer backbone, so that it mimics the chemical composition of ECM glycosaminoglycans. In light of its suitable biodegradability, biocompatibility, immunological, antibacterial and wound-healing properties as well as good mechanical and film-forming properties [21, 22], CT has therefore been widely applied in the biomedical field as a wound dressing [23], haemostat [24] and scaffold for tissue engineering [25, 26], and as controlled drug [27] and gene [28] delivery vehicles. At the same time, CT's polycationic nature should be carefully considered for the successful formulation of bespoke drug delivery systems, due to the potential electrostatic repulsion of the polymer backbone with positively-charged growth factors, such as BMP-2 or FGF-2 [29]. Furthermore, a non-controllable electrostatic complexation of the material surface with cells may be observed following protonation of CT amino groups *in vitro* or *in vivo*, ultimately leading to cytotoxic effects. To circumvent these issues, chemical functionalisation of CT can be carried out, whereby reaction of amino and hydroxyl terminations leads to the establishment of covalent, neutralised, net-points, so that the polycationic character of CT is controlled [30, 31, 32]. CT has been carboxymethylated [33, 34], grafted with phenylalanine [35] and crosslinked with low-molecular weight segments [36], leading to a broad range of polymers. However, despite such enormous polymer variability, the fact that CT solubility is restricted to acidic conditions imposes several constraints in terms of accomplishing selective and tuenable functionalisation, so that systematic changes in molecular organisation are highly challenging. Indeed, reacted products might be unstable at acidic pH, potentially resulting in occurrence of side reactions and reduced reaction yield, so that material properties and drug loading/release functionalities cannot be systematically varied.

---

[†] Hydrogel incubation was carried out with 5 μg·mL$^{-1}$ FGF-2 solution (9 ng FGF-2 loaded per hydrogel μL; 200 μL FGF-2 solution applied per hydrogel cm$^2$; hydrogel thickness: 50 μm).

This work aimed at establishing a novel synthetic approach for the formation of bioactive CT-based systems displaying tunable network architecture and superior drug-loading functionality. It was hypothesised that selective CT functionalisation could be accomplished in aqueous systems with bifunctional segments of varied molecular weight, backbone rigidity, wettability and electrostatic charge, so that bespoke hydrogels with defined macroscopic properties could be formed. Based on the selected network architecture, it was hypothesised that hydrogel drug-loading functionality could result from the electrostatic complexation of the network chains with systematically-varied model drugs. To reach this goal, L(+)-tartaric acid (TA), poly(ethylene glycol) bis(carboxymethyl) ether (PEG), 4-Phenylenediacetic acid (4Ph), and 5-Sulfoisophthalic acid monosodium salt (PhS) were selected as dicarboxylic acids for CT functionalisation. TA and PEG have been previously applied to CT for the design of pH-responsive nanoparticles [32] and drug delivery hydrogels [36]. Here, both compounds were employed as flexible, aliphatic, neutral crosslinkers of varied molecular weight. 4Ph and PhS were selected as rigid, aromatic segments, whereby the only difference between the two molecules was the presence of a sulfonic acid group in PhS benzene ring, aimed at mimicking the growth factor-binding sites of heparin *in vivo*. Consequently, incubation in solutions containing either allura red (*AR*) as doubly negatively-charged, methyl orange (*MO*) as negatively-charged and methylene blue (*MB*) as positively-charged, model drugs was carried out in order to explore hydrogel loading functionality [37]. Thus, CT systems with systematically adjusted material properties and loading efficiencies via selective changes in network architecture were expected to be established. Ongoing research is focusing on the evaluation of the controlled drug release functionality in the presented chitosan systems.

## 2. Materials and methods

### 2.1. Materials

Low molecular weight (50000-190000 Da) CT, *N*-(3-Dimethylaminopropyl)-*N*-ethylcarbodiimide hydrochloride (EDC), *N*-hydroxysuccinimide (NHS), PEG (average $M_n \sim$ 600 Da) and hydrochloric acid solution (37%) were purchased from Sigma Aldrich (Japan). 1-Hydroxybenzotriazole (HOBt) was purchased from Kishida Chemical Co., Ltd. (Japan). PhS, 4Ph and TA were purchased from Tokyo Chemical Industry Co., Ltd. (Japan). Allura red, methylene blue and methylene orange were purchased from Tokyo Chemical Industry Co. Ltd. (Japan), Kishida Chemical Co., Ltd. (Japan) and Nacalai Tesque, Inc. (Japan), respectively. Deuterium oxide was purchased from Wako Pure Chemical Industries, Ltd. (Japan).

### 2.2. Formation of CT -based hydrogels

CT -based hydrogels were prepared by dissolving CT (0.15 g) in a HOBt (0.12 g)-water (4.23 g) solution [38]. Once the CT solution was obtained, an equimolar content of dicarboxylic acid (either TA, PEG, 4Ph or PhS) with respect to the CT amino functions was activated in water (0 °C, pH 4, 30 min.) with EDC (1 or 2.3 M) in the presence of NHS. A three-fold excess of EDC/NHS with respect to the dicarboxylic acid molar content was introduced. NHS-activated dicarboxylic acid mixture was mixed with previously-obtained CT solution and incubated at room temperature under gentle shaking, in order to allow for the nucleophilic addition reaction of CT amino groups to the crosslinker carboxylic functions to occur. Complete gel formation was observed following 1-hour reaction of CT with selected NHS-activated dicarboxylic acids. Resulting hydrogels were thoroughly washed with distilled water, followed by vacuum-drying at room temperature.

**2.3. Chemical characterisation of hydrogel networks**

Attenuated Total Reflectance Fourier-Transform Infrared (ATR FT-IR) was carried out on dry samples using a Perkin-Elmer Spectrum 100 FT-IR spectrophotometer with diamond ATR attachment. Scans were conducted from 4000 to 400 $cm^{-1}$ with 16 repetitions averaged for each spectrum at 4 $cm^{-1}$ resolution. In order to investigate the molar content of introduced crosslinkers, 5 mg hydrogel CT-PhS was exemplarily incubated in 5 M HCl solution at 60 °C until complete network degradation. Partially-degraded oligomers were dried at 40 °C with reduced pressure. $^1$H-NMR spectra were recorded at room temperature on a JEOL GSX 400 spectrophotometer (400 MHz) by dissolving 10 mg of dry oligomers in 1 mL deuterium oxide. Signals at $\delta$= 8.4 ppm and $\delta$=8.5 were selected to identify PhS species, while the signal at $\delta$=4.4 ppm was chosen to describe CT backbone. Consequently, the sulfonic acid content incorporated in the hydrogel was quantified according to the following equation:

$$S(mol.-\%) = \frac{\int H_{a,b}/3}{\int H_1} \times 100 \qquad \textbf{(Equation 1)}$$

whereby *S(mol.-%)* identifies hydrogel sulfonic acid molar content, while $\int H_{a,b}/3$ and $\int H_1$ indicates $^1$H-NMR integrations related to PhS- and CT -based signals, respectively.

## 2.4. Swelling tests

Swelling tests were carried out by incubating dry samples in 5 mL aqueous medium, either distilled water or PBS (pH 7.4), at room temperature for 24 hours. Water-equilibrated samples were retrieved, paper-blotted and weighed. The weight-based swelling ratio (*SR*) was calculated as:

$$SR = \frac{m_s - m_d}{m_d} \times 100 \quad \textbf{(Equation 2)}$$

where $m_s$ and $m_d$ are swollen and dry sample weights, respectively. Three replicas were used for each sample composition, so that *SR* results were expressed as average ± standard deviation. Single factor ANOVA was carried out to determine the statistical significance between experimental groups. A value of $p < 0.05$ was considered to be statistically significant.

## 2.5. Compression tests

Water-equilibrated hydrogel discs (ø 0.8 cm) were compressed at room temperature with a compression rate of 3 mm·min$^{-1}$ (EZ test, Shimadzu Corporation, Japan). A 500 N load cell was operated up to sample break. The maximal compressive stress ($\sigma_{max}$) and compression at break ($\varepsilon_b$) were recorded, so that the compressive modulus (*E*) was calculated by fitting the linear region of the stress-strain curve. At least three replicas were employed for each composition and results expressed as average ± standard deviation.

## 2.6. Thermogravimetric analysis

Thermogravimetric analysis (TGA, TG/DTA 6200, Seiko Instruments Inc., Japan) was conducted on dry samples in order to investigate the thermal stability of formed networks. TGA tests were carried out under nitrogen atmosphere (50 mL·min$^{-1}$ nitrogen flow rate range) with 10 °C·min$^{-1}$ heating rate from 10 to 500 °C.

## 2.7. Scanning electron microscopy

The internal architecture of formed hydrogels was investigated following freeze-drying. Freeze-dried samples were mounted on metal stubs using double-sided carbon adhesive tape and coated with osmium tetroxide by HPC-30 plasma coater (Vacuum Device Inc.). Coated samples were inspected with JEOL JSM-6700FE field emission scanning electron microscope, whereby the pore size was determined by direct measurements (n= 2-8).

## 2.8. Drug loading into CT hydrogel

Allura red AC (*AR*, $\lambda_{max}$ = 504 nm), methyl orange (*MO*, $\lambda_{max}$ = 505 nm) and methylene blue (*MB*, $\lambda_{max}$ = 661 nm) were used as doubly negatively-, negatively- and positively-charged model drugs, respectively, in order to estimate hydrogel loading functionality. A stock solution of each dye was prepared in distilled water at a final concentration of 0.1 mg·mL$^{-1}$. The stock solution was further diluted before incubation of a water-equilibrated hydrogel (10-50 mg) in 4 mL dye solution (0.025 mg·mL$^{-1}$). At selected incubation time points, the incubating solution was collected in a cuvette and analyzed via a UV spectrophotometer U-3010 (Hitachi High-Technologies Corporation, Japan). Thus, hydrogel loading with model drugs was quantified according to the following equation:

$$XX\ loading = \left(1 - \frac{A_{max,d} - A_{max,t}}{A_{max,d}}\right) \cdot 100 \quad \textbf{(Equation 3)}$$

whereby *XX* identifies the loading model drug, i.e. *AR*, *MO* or *MB*; $A_{max,d}$ indicates the maximum absorbance of the model drug solution before incubation of the hydrogel sample, while $A_{max,t}$ describes the maximum absorbance of the solution supernatant following hydrogel incubation at selected loading time points.

## 3. Results and discussion

CT hydrogels with varied crosslinkers, controlled macroscopic properties and tunable drug-loading functionality were successfully synthesised (Scheme 1). These hydrogels were colourless and transparent in their gelled unloaded state; following drug loading the gels took on the colour of the relevant drug (Figure S1, Supporting Information shows an image of hydrogel CT-PhS following AR loading). CT was dissolved in HOBt aqueous solution, as an effective, reliable and cell-friendly system to directly functionalise CT avoiding any pre-derivatisation step or acidic conditions [38]. Reaction with selected NHS-activated dicarboxylic acids led to complete gel formation within 30 min reaction. Here, nucleophilic addition of CT amino and hydroxyl functions to activated crosslinkers takes place [22], leading to a covalent network consisting of hydrolytically-cleavable amide net-points (Scheme 1, A). Thus, either intra- or inter-molecular crosslinks can be introduced in CT chains, depending on the molecular rigidity, molecular weight and feed ratio of selected crosslinking segments (Scheme 1, B). Consequently, materials with varied macroscopic properties and internal geometry could be accomplished. Furthermore, hydrogel electrostatic charge was expected to be tuned based on resulting network architectures, so that loading with a wide range of drugs could be successfully obtained via electrostatic drug-hydrogel complexation (Scheme 1, C). In the following, CT-based hydrogels will be characterised as for their network architecture, swellability, thermo-mechanical properties, material morphology and drug-loading functionality, aiming at establishing systems with defined structure-property-function relationships. Samples are coded as CT(XX%)-YYY, whereby CT indicates a CT-based hydrogel, XX identifies the wt.-% CT concentration in the reacting mixture, while YYY states the type of crosslinker introduced in resulting networks, either TA, PEG, 4Ph or PhS.

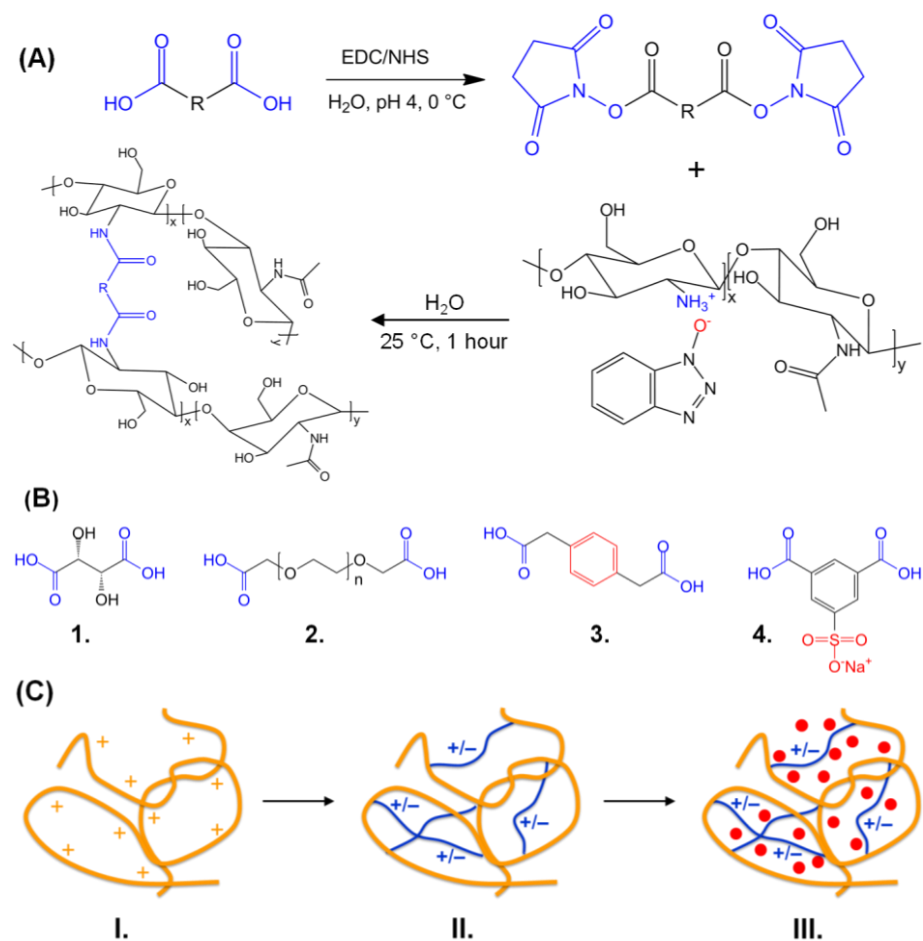

**Scheme 1.** Formation of CT-based hydrogels with varied network architecture and tunable drug-loading functionality. (A): A dicarboxylic acid segment is activated with EDC/NHS and reacted with an aqueous solution of CT-HOBt. (B): Molecular formula of selected bifunctional segments: Tartaric acid (1., TA), poly(ethylene glycol) bis(carboxymethyl) ether (2., PEG), 1,4-Phenylenediacetic acid (3., 4Ph), Monosodium 5-Sulfoisophthalate (4., PhS). (C): Drug-loading functionality of CT-based hydrogels: native CT (−) behaves as a polycation in aqueous solution due to the protonation of amino and hydroxyl functions (I). By covalently functionalising CT with either negatively- or positively-charged bifunctional segments (−), hydrogel electrostatic charge can be promptly varied (II). Adjustment in backbone electrostatic charge is exploited to achieve hydrogel loading with drug model compounds (●) via electrostatic complexation (III).

### 3.1. Chemical composition and network architecture

Vacuum-dried CT-based networks were investigated via ATR FT-IR in order to explore the chemical composition in formed materials and elucidate the mechanism of network formation. Spectral analysis was then combined with $^1$H-NMR spectroscopy on oligomers resulting from network degradation in acidic conditions, in order to quantify the crosslinker content in the covalent networks. Figure 1 displays FT-IR spectra of native and crosslinked CT following reaction with selected bifunctional segments. Distinctive FT-IR bands of CT are at 3450 cm$^{-1}$ (O–H stretch), 2872 cm$^{-1}$ (C–H stretch), 1646 cm$^{-1}$ (Amide I), 1590 cm$^{-1}$ (Amide II), 1152 cm$^{-1}$ (bridge-O stretch), and 1090 cm$^{-1}$ (C–O stretch) [34, 36]. Each of these bands was found in all acquired spectra, confirming the CT-based composition in resulting samples. At the same time, increased intensities of Amide I and II bands, while an additional band at 1740 cm$^{-1}$, were observed in crosslinked, in comparison to native CT, spectra. The former finding is likely to confirm the presence of a covalent network consisting of amide net-points formed following the reaction between CT amino terminations and activated carboxylic groups of the crosslinkers. Furthermore, the new peak at 1740 cm$^{-1}$ is likely to identify the C=O band of ester groups formed following the reaction of hydroxyl functions with activated diacids.

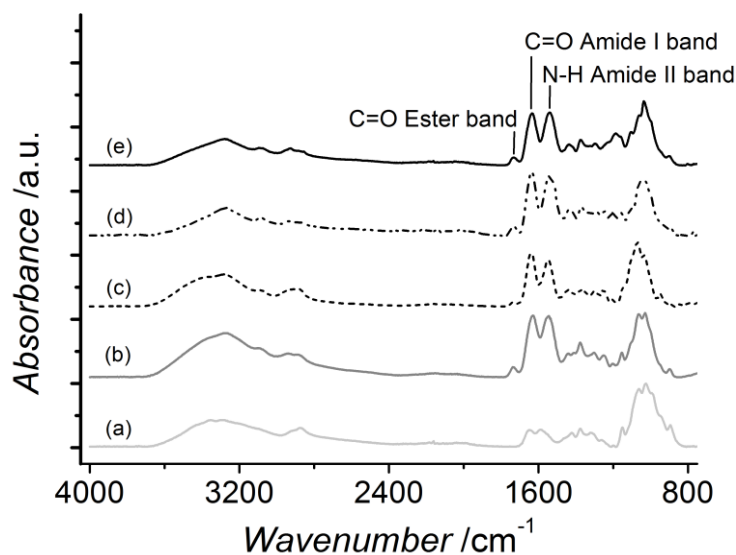

**Figure 1.** Exemplary ATR-FTIR spectra of CT (a), CT(2.2%)-TA (b), CT(2.2%)-PEG (c), CT(2.2%)-4Ph (d), and CT(2.2%)-PhS (e). The presence of Amide I (1650 cm$^{-1}$), Amide II (1550 cm$^{-1}$) and Ester (1730 cm$^{-1}$) bands in functionalised, compared to native, CT confirms the formation of a covalent network via reaction of carboxylic terminations of selected bifunctional segments with amino and hydroxyl functions of CT, respectively.

The lower intensity of this band compared to amide bands suggests that hydroxyl group functionalisation is less favoured, which is expected due to the lower reactivity of hydroxyl compared to amino functions. Other than that, no additional band characteristic of selected diacids was observed in resulting sample spectra, likely due to the much lower crosslinker, compared to CT, content.

Besides ATR FT-IR, it was of interest to quantify the crosslinker content effectively incorporated in formed hydrogels. This information was crucial in order to elucidate the molecular architecture of resulting networks and understand how this affected hydrogel macroscopic properties and drug-loading functionality. Covalent, non-soluble, hydrogel networks are indeed challenging to characterise precisely, since main molecular parameters, such as network crosslinking density, can only be determined indirectly following equilibrium-swelling and rubbery-elasticity theories. However, noisy experimental conditions are usually associated with the determination of these molecular parameters, so that only semi-quantitative conclusions can be drawn [39]. In order to overcome these limitations, a reverse approach was pursued for the characterisation of network architecture, whereby selected a CT hydrogel was fully-degraded in acidic conditions and resulting oligomers analysed via $^1$H-NMR spectroscopy.

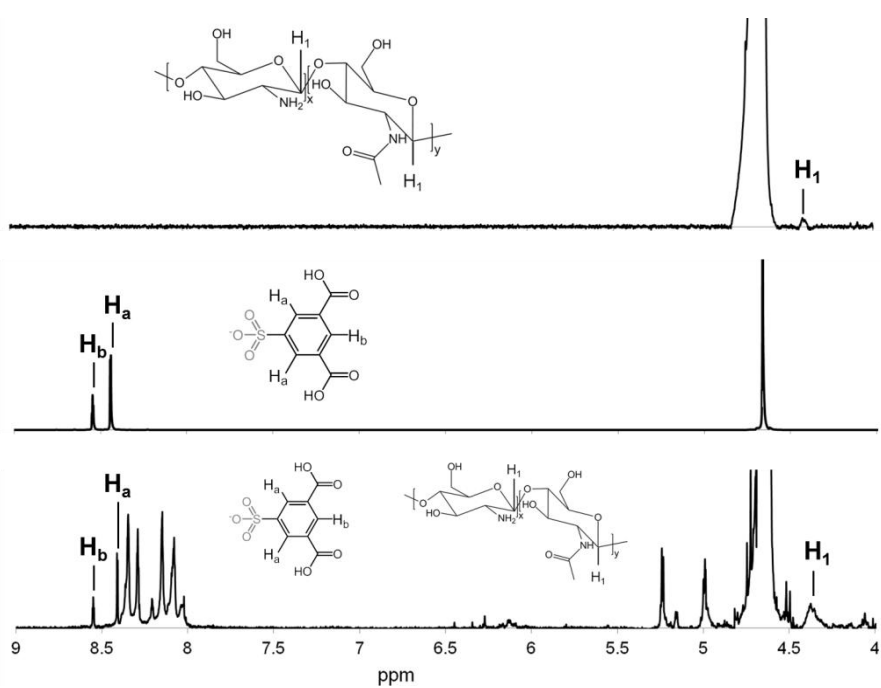

**Figure 2.** $^1$H-NMR spectra (D$_2$O, 25 °C) of native CT (top), PhS (middle) and partially-degraded oligomers (bottom) formed following hydrolysis (5 M HCl, 60 °C) of hydrogel CT(2.2%)-PhS. Signals related to PhS- (H$_a$, H$_b$) and CT- (H$_1$) based species were clearly identified in oligomers spectra, so that the PhS content in resulting hydrogels was successfully quantified (~ 18 mol.-%).

Accordingly, samples of CT(2.2%)-PhS, as an exemplary hydrogel system, were selected since it was of particular interest to investigate how introduced sulfonic acid moieties influenced hydrogel swellability and drug-loading functionality (sulfonic acid being responsible for basic growth factor sequestration in tissue ECM).

Figure 2 displays the $^1$H-NMR spectra of native CT, PhS and partially-degraded oligomers, respectively. Here, PhS- (H$_a$, H$_b$) and CT- (H$_1$) related signals could be clearly distinguished in the spectrum of partially-degraded network oligomers, whereby no overlap with any other species was observed. This finding provides supporting evidence that network degradation only takes place via hydrolytic cleavage of either amide or ester bond net-points, resulting from the PhS-mediated CT functionalisation. Consequently, CT-related signal H$_1$ could still be identified following network degradation, so that quantitative information was successfully obtained via integration ratio of PhS-related signals (H$_a$, H$_b$) and CT reference signal (H$_1$, Equation 1). Accordingly, a sulfonic acid content of 18 mol.-% was determined. As for comparison, Muzzarelli functionalised CT with 5-formyl-2-furansulfonic acid to obtain an anticoagulant polymer analogue, whereby a sulfation degree of less than 4% was obtained, although an additional chloromethylation step was required [40]. Following a similar approach, Hoven et al. synthesised *N*-sulfofurfuryl-bearing CT films [41], so that the sulfation degree was increased up to nearly 8%; however, the reaction was carried out in the bulk state so that a non-homogeneous functionalisation was expected. In contrast to previous approaches, requiring additional reaction steps and resulting in low, non-tunable functionalisation, this synthetic route is highly advantageous, since it enables the direct functionalisation of CT with enhanced degree of functionalisation, so that functional, tunable hydrogels can be successfully obtained. Due to the versatility of this synthesis, it is expected that the degree of functionalisation can be increased further by tuning the crosslinker: CT feed molar ratio.

## 3.2. Swellability of CT-based hydrogels

The swelling ratio (*SR*) was quantified in either distilled water (pH 6.5) or PBS (0.1 M, pH 7.4) in order to investigate material behaviour in physiologically-relevant conditions. Figure 3 describes *SR* values for the different hydrogel compositions; *SR* of resulting chitosan networks was varied in the range of 300-1100 wt.-%; networks crosslinked with PEG showed the highest *SR* variation (*SR*: 500-1100 wt.-%), followed by samples CT-TA (300-600 wt.-%) and CT-PhS (400-600 wt.-%), whilst samples CT-4Ph took up the lowest amount of aqueous medium (300-400 wt.-%). Hydrogels CT-PhS displayed slightly higher *SR* in PBS compared to distilled water, although the opposite trend was observed in all other networks, with significant differences observed in systems CT(1.5%)-TA and CT(2.2%)-PEG. Moreover, changes in reacting CT concentration significantly affected the swellability of PEG- (following incubation in both distilled water and PBS) and TA-based (following incubation in PBS only) systems, despite there being only a very small change in reacting CT concentration (1.5-2.2 wt.-%). Overall, the following parameters were observed to rule the swelling behaviour of resulting CT networks: (i) network architecture, whereby variations were mainly based on the wettability, backbone rigidity and molecular weight of applied crosslinkers; (ii) CT concentration in the crosslinking mixture; (iii) electrostatic contributions resulting from the swelling medium, occurrence of free (i.e. amino) CT terminations and/or ionically-charged groups (i.e. sulfonic acid moieties) introduced following crosslinking.

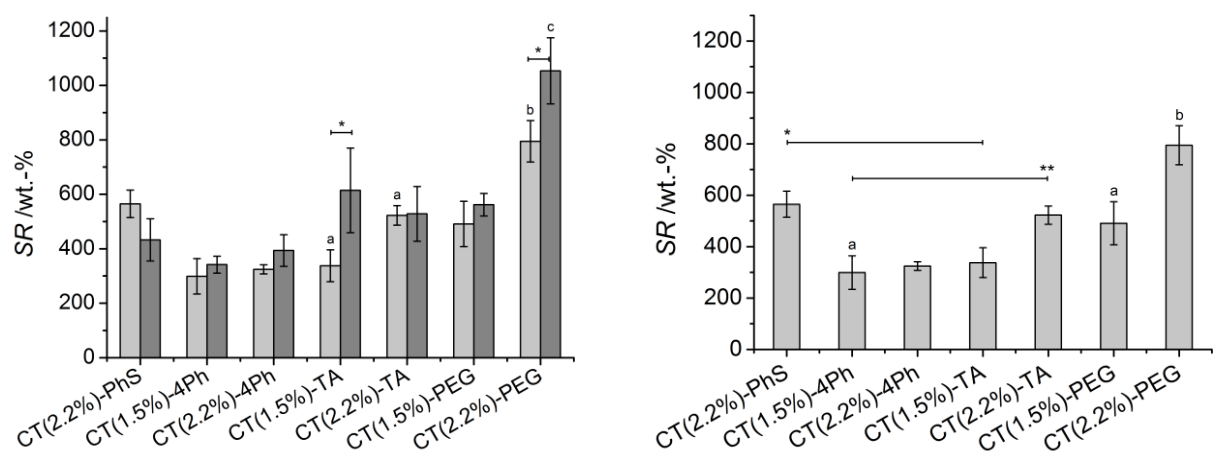

**Figure 3.** Left: Swelling ratio (*SR*) of CT-based hydrogels following equilibration in either distilled water (grey) or PBS (light grey) at room temperature. '*' and 'a' indicate that related mean differences are significant at 0.05 level (Bonferroni test). 'b' and 'c' indicate that corresponding mean values are significantly different to all mean values described in the plot (at 0.05 level, Bonferroni test). Right: *SR* of CT-based hydrogels following equilibration in PBS at room temperature. '*' and '**' indicate that corresponding mean values are significantly higher compared to all other mean values described below the black line (at 0.05 level, Bonferroni test); 'a' indicates that compared mean values are significantly different between each other at 0.05 level (Bonferroni test); 'b' indicates that corresponding mean value is significantly different from all other mean values described in the plot (at 0.05 level, Bonferroni test).

Hydrogel swelling is driven by the spontaneous mixing of network chains with water, whilst the presence of covalent net-points among polymer chains prevents the dissolution of the network, although contribution from ionic moieties also plays a major role. In order to understand the swelling behaviour of obtained CT systems, it is therefore crucial to focus on their molecular organisation. Given that the same reaction mechanism was employed for network formation (Scheme 1, A) and a constant molar feed ratio was applied, differences in network crosslinking density are unlikely to be expected among the different CT systems. Consequently, the network architecture will be mainly affected by the molecular structure of each crosslinking segment, described by its molecular weight, molecular rigidity, wettability and electrostatic charge, although the ionic contribution deriving from non-reacted amino terminations of chitosan also needs to be considered. Considering the former aspect, the enhanced swellability observed in PEG- (in either distilled water or PBS), and in some cases in TA- (mainly in PBS), based systems can be mainly ascribed to the increased hydrophilicity and flexibility of both PEG and TA compared to employed aromatic, rigid crosslinkers, such as 4Ph and PhS. Likewise, the lower swellability observed in TA- compared to PEG-based systems is likely related to the varied crosslinker molecular weight, since this molecular parameter rules network elasticity and extensibility during swelling [42]. Thus, higher swelling is likely to be expected in networks crosslinked with longer segments compared to networks crosslinked with segments of decreased molecular weight. Other than crosslinker molecular weight and backbone rigidity, the swelling medium is also shown to play a major role in hydrogel swellability. The reason for this interesting behaviour in obtained CT systems is to be found in the polycationic nature of native CT. CT $pK_a$ is 6.2-7 [21, 32], so that its amino terminations are protonated in distilled water (pH 6.5), while they are likely neutralised in PBS (0.1 M, pH 7.4). This effect is reduced in crosslinked CT and inversely-related to the network crosslinking density, since free amino groups will be functionalised to form amide bonds, as observed via ATR FT-IR (Figure 1). Consequently, in CT networks crosslinked with non-ionically charged segments, i.e. TA, PEG, and 4Ph, non-functionalised amino groups will be protonated in distilled water, so that the electrostatic repulsion between them will promote increased swelling. In contrast, no repulsion will be present in PBS since all free amino terminations will be neutralised, resulting in decreased *SR*. This behaviour is slightly reversed in the case of PhS-functionalised CT, given that additional, negatively-charged moieties are established along crosslinked chains. In distilled water, sulfonic acid groups will be able to mediate additional electrostatic crosslinks with free, protonated amino groups, so that decreased *SR* is observed (Figure 3). At the same time, such electrostatic interactions will disappear in PBS, since free CT amino groups will be neutralised in these conditions. Consequently, *SR* is slightly increased due to the electrostatic repulsion between negatively-charged sulfonic acid moieties.

### 3.3. Thermo-mechanical properties of CT systems

Thermal analysis was carried out via TGA on dry networks in order to get an insight on the thermal material stability and network morphology, while mechanical properties were investigated via compression tests on water-equilibrated hydrogels. Figure 4 describes TGA thermograms of native and crosslinked CT samples. Thermal phenomena such as water loss, material decomposition and combustion were exhibited by all samples. At the same time, no additional thermal event was observed in crosslinked compared to native CT, indicating that no presence of by-products or non-reacted moieties could be observed. The synthesis of pure, fully-crosslinked covalent networks was therefore proved following hydrogel formation and washing. Native CT displayed a weight loss between 50 and 150 °C which is obviously related to the water evaporation from the sample. Such weight loss is also observed in the crosslinked materials although in a smaller extent, especially in the case of TA-based networks, probably due to the presence of covalent net-points limiting the amount of fluid imbibed and retained by the network. Compared to native CT, crosslinked samples displayed lower thermal stability (< 250 °C). These observations likely suggest that inter- rather than intramolecular functionalisation of polymer chains is obtained following network formation, so that CT crystallisation is inhibited [21, 35]. Furthermore, the material thermal stability is affected only slightly by the molecular rigidity of introduced aromatic or aliphatic crosslinkers.

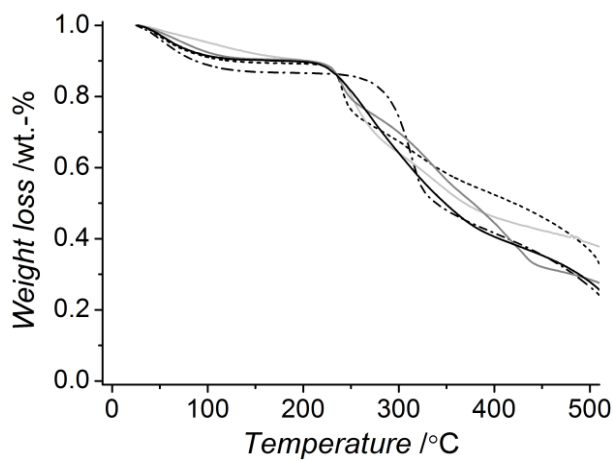

**Figure 4.** TGA thermograms of CT networks: CT (–··–), CT(2.2%)-TA (—), CT(2.2%)-PEG (—), CT(2.2%)-4Ph (—), CT(2.2%)-PhS (···).

Water-equilibrated hydrogels were investigated by compression tests at room temperature. Here, stress-strain curves were varied based on the network architecture of hydrogel samples (Figure 5), so that resulting compressive moduli ($E$) were obtained in the low kPa range (2-9 kPa). PEG- and 4Ph-based systems showed the lowest and highest $E$ value, respectively, while mechanical properties of TA- and PhS-based systems were within the two extremes. These mechanical observations are in agreement with aforementioned network architecture and swelling considerations, whereby samples showing increased $SR$ exhibited decreased $E$ and vice-versa. PEG is a polymeric hydrophilic, aliphatic segment, resulting in systems with enhanced elasticity and swellability, and reduced compressive modulus, in comparison to the other CT hydrogels. In contrast, 4Ph is an aromatic, bulky crosslinker, thereby leading to the formation of hydrogels with increased mechanical properties and decreased swellability. TA and PhS display intermediate molecular features between these two compounds. TA is a low-molecular weight aliphatic segment, so that the molecular weight between two covalent netpoints will be reduced in TA- compared to PEG-based systems. Consequently, decreased elasticity and swellability as well as increased compressive modulus will be expected in resulting networks. PhS is an aromatic crosslinker (unlike PEG molecule) containing an additional ionically-charged sulfonic acid moiety with respect to 4Ph compound.

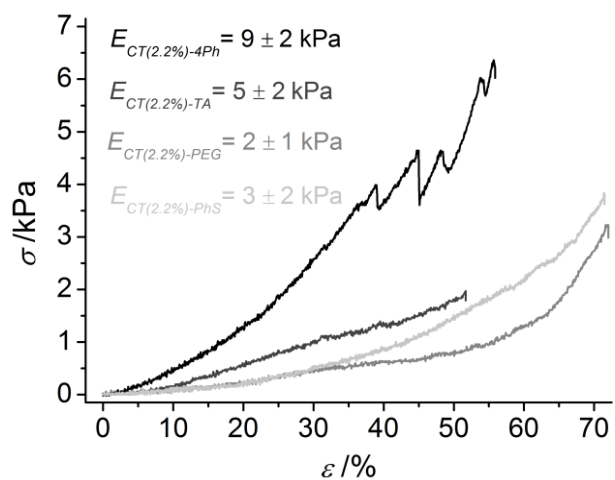

**Figure 5.** Compressive stress (σ)-strain (ε) curves of chitosan-based hydrogels. From top to bottom: CT(2.2%)-4Ph (black), CT(2.2%)-TA (dark gray), CT(2.2%)-PEG (gray), , CT(2.2%)-PhS (light gray). Compressive moduli were calculated (n ≥ 3) in the linear region and were observed to be affected by the specific network architecture.

The decreased compression modulus observed in CT-PhS compared to CT-4Ph samples is therefore likely related to the electrostatic repulsion occurring during the compression of the hydrogel network. Here, crosslinked chains of CT-PhS network become closer, since the material is stressed in compression. This molecular mechanism is likely to explain the decreased compression modulus in CT-PhS with respect to CT-4Ph samples, although the swelling ratio in distilled water between these sample formulations is similar.

### 3.4. SEM morphological investigation

Since obtained CT systems were intended for biomedical applications, e.g. as drug reservoirs or scaffold for tissue engineering, it was of interest to explore material surface morphology and inner geometry. In both applications, the presence of pores play an important role in hydrogel properties, such as swelling [21], compressability [43] and cell adhesion [44], and can enhance nutrient diffusion, waste exchange as well as induce local angiogenesis following material implantation *in vivo* [25]. Figure 6 displays SEM pictures of freeze-dried network surfaces and cross-sections. Samples exhibited a porous morphology, whereby pore size (Ø 60-400 μm) was changed based on the selected CT system. TA-crosslinked CT exhibited a highly porous architecture (Ø 271±100 μm) with nearly-rounded pores observed either in the outer or inner sections, in contrast to hydrogels CT-PEG (Ø 82±24 μm) and CT-4Ph (Ø 62±5 μm).

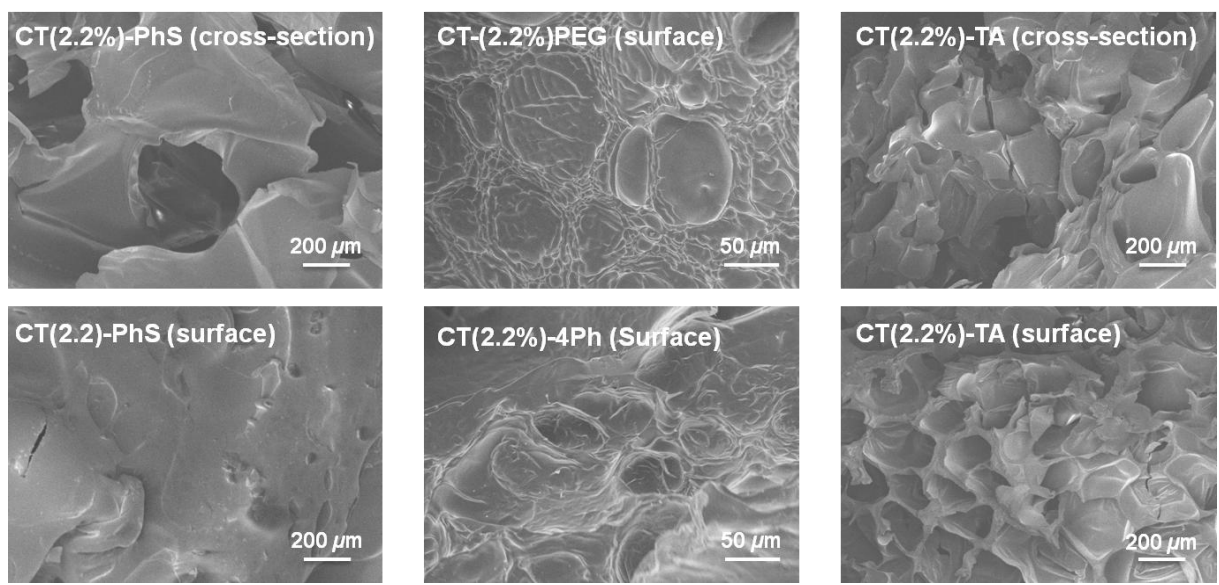

**Figure 6.** Scanning electron microscopy (SEM) on freeze-dried CT-based networks.

At the same time, PhS-based samples revealed the presence of bigger pores (Ø 411±47 µm) in the inner structure, while the material surface was nearly bulky. Interestingly, the pore size of freeze-dried systems was that of previously-reported, gas-foamed CT-based scaffolds [25]. Given that no specific scaffold formation method was applied, it is likely that observed pores derived from the reaction conditions employed during hydrogel formation. Air-bubbles can indeed be introduced in the polymer solution during stirring, remaining entrapped in the hydrogel following network formation. Consequently, pores are expected to be formed in the resulting material following freeze-drying [45].

Selected crosslinkers are likely to influence air-bubble stabilisation and thereby resulting material morphology. Slightly-hydrophobic moieties are supposed to enhance phase separation in foamed-like solutions [46], so that homogeneous pores of decreased size can be formed, as observed in the case of 4Ph-based systems. In contrast, crosslinkers of increased hydrophilicity are expected to destabilise the air-liquid phase separation, so that coalescence and disproportionation of air-bubble is likely, ultimately resulting in pores of increased size and decreased occurrence. Such morphological features are clearly observed in samples CT-TA and, to some extent, in samples CT-PhS, but not in the case of samples CT-PEG. The varied molecular weight between TA and PEG is likely to count for the different material morphology between the two systems, since polymers of increased molecular weight will have an effect on the solution viscosity, which is another physical parameter crucial for air bubble stabilisation [46]. With the presented synthetic strategy, it was therefore possible to accomplish CT networks with crosslinking segments of varied wettability, so that pores with varied pore size could be formed adjusted towards specific clinical applications, such as osteoid ingrowth (Ø 40-100 μm) or bone regeneration (Ø 100-350 μm) [25].

### 3.5. Drug-loading functionality in CT systems

In view of the versatility of these CT systems, it was of particular interest to investigate whether the formation of selected network architectures could result in advanced and tailorable drug-loading functionalities. By functionalisation of CT with selected crosslinkers, it was hypothesised that network electrostatic charge could be systematically tuned in order to promote hydrogel-drug electrostatic complexation with a wide range of drugs. Hydrogels CT-PEG, CT-4Ph and CT-PhS were incubated in aqueous solutions of either *AR*, *MO* or *MB* and the maximum absorbance of the supernatant monitored over time, in order to quantify any drug loading onto the material (Equation 3). Figure 7 (a) displays absorbance curves of *MB* solution at different time points following incubation with hydrogel CT-PhS. Here, the maximum absorbance peak is decreased at increased incubation time points, indicating a reduction of *MB* content in the supernatant, likely related to *MB* loading into the hydrogel. Significant decrease in absorbance was observed following 24 hours incubation, corresponding to the time window nedeed to enable equilibration of CT networks with distilled water.

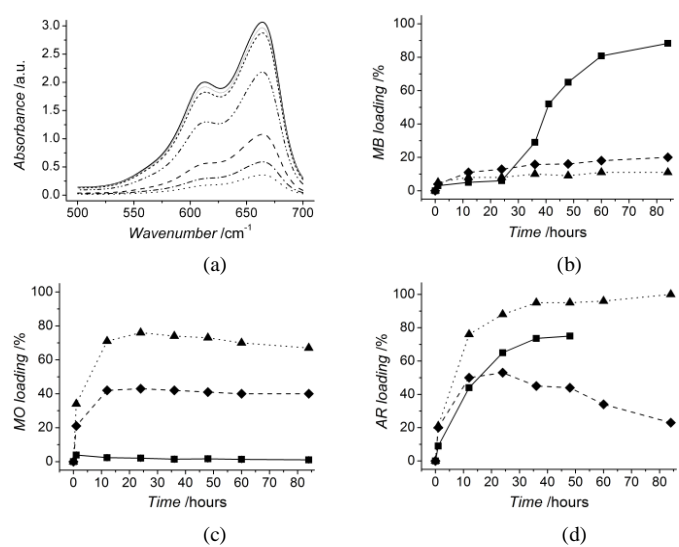

**Figure 7.** Typical loading behaviour of CT-based systems studied via UV-vis incubation experiments with dyes of varied electrostatic charge. Absorbance curve of MB-containing solution during incubation with hydrogels CT(2%)-PhS (a). Peak of MB absorbance at 661 nm is decreased at increased incubation time points (0, 12, 24, 36, 48, 60, 84 h), providing evidence of MB loading in the CT hydrogel. MB (b), MO (c) and AR (d) loading

curves in hydrogels CT(2%)-PhS (–■–), CT(2%)-4Ph (--♦--), CT(2%)-PEG (··▲··).Triplicates were applied for the loading of hydrogels CT(2%)-PhS with MO; minimal standard deviation was observed between replicas at each loading time point (*Loading*: 1.04 ± 0.74 – 3.89 ± 1.07), giving supporting evidence of the system reproducibility. Hydrogels are selectively loaded with dyes of opposite electrostatic charge, due to the electrostatic hydrogel-dye complexation occurring following hydrogel incubation.

At the same time, the maximum absorbance peak was mostly suppressed following 84 hours hydrogel incubation, indicating that *MB* loading onto the hydrogel was almost complete. From a molecular point of view, *MB* is a positively-charged compound, similarly to growth factors such as FGF-2 or BMP-2. Due to its cationic character, electrostatic repulsion with CT, resulting in minimal loading efficiency, should be expected [27, 28]. *MB* was dissolved in distilled water, at which pH protonation of CT amino groups takes place, as demonstrated via swelling tests (Figure 3). The only reason explaining the high *MB* loading onto hydrogels CT-PhS is therefore related to the presence of sulfonic acid moieties covalently bonded to CT backbone: such negatively-charged groups are likely to mediate electrostatic complexation with *MB* positively-charged groups, so that localised electrostatic complexation can successfully occur, ultimately resulting in almost complete loading. Further to that, no visible colour change was observed when loaded hydrogels were incubated (for at least three days) in distilled water, suggesting that only minimal burst release may be expected following hydrogel loading.

In order to further understand the molecular mechanism ruling hydrogel loading capability, hydrogel loading with selected model drugs was quantified by monitoring the variation in maximum absorbance. Figure 7 (b) describes *MB* loading plots for samples CT-PhS, CT-4Ph and CT-PEG. As observed from the suppressed absrobance peak (Figure 7, b), CT-PhS could load up to 88% *MB* following 84 hours incubation, while the other two samples showed less than 20 % loading. Compared to that, the situation is completely reversed when hydrogels are incubated with solutions containing negatively-charged *MO* (Figure 7, b), whereby samples CT-PhS displayed only mininal adsorption (< 5%), while increased loading was observed in hydrogels CT-4Ph (~ 40%) and CT-PEG (~ 65%). Different loading capabilities were again observed when hydrogels were incubated with *AR* (Figure 7, c), so that samples CT-PEG highlighted more than 90% loading, followed by samples CT-PhS (~ 70%), while samples CT-4Ph displayed nearly 50% loading. The wide difference between the loading capabilities of tested hydrogels confirmed a direct relationship with the molecular architecture of selected hydrogel networks. Hydrogels CT-PhS were unable to load negatively-charged *MO*, an observation which is directly ascribed to the electrostatic repulsion of this compound with CT-bound sulfonic acid moiety. In contrast to that, significant loading was observed in the same system with doubly-negatively charged *AR*, whereby the hydrogel turned red following dye incorporation in the material (Figure S1, Supporting Information). This finding is likely related to the fact that electrostatic complexation between non-functionalised, protonated CT amino terminations is more likely in the case of *AR* compared to *MO*, due to the increased density of negatively-charged moeties. Consequently, the previously-observed drug repulsion with PhS-functionalised CT could be overcome. Other than that, the high *AR* and *MO* as well as low *MB* loadings of PhS-free CT systems further support aferomentioned explanations, thereby demonstrating that tunable and localised hydrogel-drug electrostatic complexation could be successfully established based on the systematic variation in network architecture. Therefore, the drug-loading capability of formed CT systems was successfully ruled by the polymer functionalisation with selected bifuctional crosslinkers, so that a wide range of varied model drugs could be applied. In light of this direct relationship between network architecture and drug-loading functionality, the loading efficiency was significantly enhanced compared to previously-reported alginate- [9], poly(glutamic acid)- [10], , and PEG- [5, 29] based hydrogels.

## Conclusions

Functionalised CT systems were successfully prepared. Bifunctional segments varying in molecular weight, backbone rigidity and electrostatic charge were applied to selectively functionalise CT in order to investigate the changing trends in network architecture, macroscopic properties and drug-loading capability. Systematically-adjusted swellability, and compressability were observed by applying selected crosslinkers, while a varied range of model drugs were successfully loaded into the systems via hydrogel-drug electrostatic complexation. In order to form tunable CT systems via a reliable, effective and cell-friendly synthetic route, CT was functionalised in a HOBt-water mixture, allowing for the direct formation of TA-, PEG-, 4Ph- and PhS-based hydrogels. Precise quantification of the crosslinker content was reversely obtained via acidic degradation of obtained hydrogels, so that an enhanced degree of CT functionalisation ($S\sim$ 18 mol.-%) was determined via $^1$H-NMR on exemplarily-formed oligomers CT-PhS. Network thermal stability was decreased compared to native CT, likely due to the formation of inter- rather than intramolecular netpoints, so that polymer crystallinity was decreased. At the same time, internal material morphology was affected depending on the molecular structure of incorporated crosslinkers. Based on the selected network architecture, swellability ($SR$: 299±65–1054±121 wt.-%) and compressability ($E$: 2.1±0.9–9.2±2.3 kPa) could be promptly adjusted. Furthermore, the electrostatic charge of hydrogel networks was varied leading to the promotion of selective complexation with a wide range of model drugs, i.e. *AR*, *MO* and *MB*, so that hydrogel loading was successfully adjusted (1-100%). Introduction of heparin-mimicking sulfonic acid moieties was crucial to ensure full *MB* loading in CT-PhS networks. This finding was not accomplished in any of the other network architectures, including 4Ph-bearing networks. This confirms the hypothesis that the sulfonic acid group is key in promoting chitosan binding to positively-charged model drugs, which is not possible in native chitosan. Consequently, a direct method to promptly alter the polycationic feature of chitosan was successfully established, potentially paving the way to selective and controlled loading and release of positively-charged growth factors. Because these CT systems can mimic biological tissues on a mechanical, molecular and functional level, next steps are focusing on the application of these materials as drug reservoirs, whereby sustained drug release can be obtained. Since selective loading was demonstrated with model drugs, CT systems will be investigated with growth factors, e.g. FGF-2 and BMP-2, in order to further explore the biofunctionality of these hydrogels as drug carriers.

## Acknowledgements

The authors wish to thank the Sasakawa Foundation of Great Britain for partly sponsoring this project via the Butterfield Award. This work was partially funded through WELMEC, a Centre of Excellence in Medical Engineering funded by the Wellcome Trust and EPSRC, under grant number WT 088908/Z/09/Z. The authors wish to thank O. Tomoya and W. Vickers for their assistance with SEM investigation.